\documentclass[letterpaper]{article} 
\usepackage{aaai25}  
\nocopyright
\usepackage{times}  
\usepackage{helvet}  
\usepackage{courier}  
\usepackage[hyphens]{url}  
\usepackage{graphicx} 
\usepackage{natbib}  
\usepackage{caption} 
\usepackage{algorithm}
\usepackage{algorithmic}
\usepackage{xcolor}

\usepackage{newfloat}
\usepackage{listings}
\DeclareCaptionStyle{ruled}{labelfont=normalfont,labelsep=colon,strut=off} 
\floatstyle{ruled}
\newfloat{listing}{tb}{lst}{}
\floatname{listing}{Listing}
\title{AI Agents and the Law}
\author {
    Mark O. Riedl\textsuperscript{\rm 1},
    Deven R. Desai\textsuperscript{\rm 2}
}
\affiliations {
    \textsuperscript{\rm 1}College of Computing\\
    \textsuperscript{\rm 2}Scheller College of Business\\
    Georgia Institute of Technology\\
    riedl@cc.gatech.edu, deven.desai@scheller.gatech.edu
}

\usepackage{bibentry}

\begin{document}

\maketitle

\begin{abstract}
As AI becomes more ``agentic,'' it faces technical and socio-legal issues it must address if it is to fulfill its promise of increased economic productivity and efficiency.
This paper uses technical and legal perspectives to explain how things change when AI systems start being able to directly execute tasks on behalf of a user.
We show how technical conceptions of agents track some, but not all, socio-legal conceptions of agency. That is, both computer science and the law recognize the problems of under-specification for an agent, and both disciplines have robust conceptions of how to address ensuring an agent does what the programmer,  
or in the law, the principal desires and no more. 
However, to date, computer science has under-theorized issues related to questions of loyalty and to third parties that interact with an agent, both of which are central parts of the law of agency.
First, we examine the correlations between implied authority in agency law and the principle of value-alignment in AI, wherein AI systems must operate under imperfect objective specification.
Second, we reveal gaps in the current computer science view of agents pertaining to the legal concepts of disclosure and loyalty, and how failure to account for them can result in unintended effects in AI ecommerce agents.
In surfacing these gaps, we show a path forward for responsible AI agent development and deployment.

\end{abstract}

\section{Introduction}

As Artificial Intelligence becomes more “agentic,” it faces technical and socio-legal issues that must be addressed if it is to fulfill its promise of increased economic productivity and efficiency. This paper explains what changes from a technical and legal perspective when an AI system moves beyond relatively passive outputs, such as text or images, to active execution of tasks. By examining the technical issues and socio-legal issues, we show how technical conceptions of agents track some, but not all, conceptions of agency according to the law. By revealing the gaps in the current computer science view of agents, we add to core ideas in value alignment and show a path forward for responsible AI Agents.

The Large Language Models (LLMs) that underpin many of the most prominent, commercially oriented artificial intelligence systems receive textual information in the form of a prompt and produce text as output. 
While LLMs are highly fluent language generators, they have also been trained on large amounts of text that describe everyday tasks such as product reviews, travel itineraries, and people describing how they have solved math and science problems.  
As such, LLMs are often capable of explaining or instructing people on how to accomplish a task. 
LLMs, in their base configuration, are {\em agentic} in the simplest meaning of the term;
just as one might ask a human for an itinerary for a given trip, an LLM can create your trip's itinerary or other passive lists of actions or plans depending on your prompt. 
But, LLMs cannot natively act directly on the world and respond to the way the world changes as a result.

Even when an LLM-based AI system can only produce text or an image, the theoretical and actual harms that can be caused are well documented, including, but not limited to, hallucinations of facts, prejudicial biases, defamation, and affecting mental health and well-being~\cite{Wolf2017ShouldHave,Weidinger2021EthicalAS,Gabriel2024TheEO}.
Although these harms are important, they are blunted by the relative passivity of the systems. 
It takes a human user to interpret and act on the LLM's output. 

Once an AI system has the ability to act in a way that can change the state of the world,
a broader set of concerns emerges. 
In simple terms, the core technical and socio-legal issues around agents---how to ensure that agents do what we want and not cause harm---becomes acute. 
This paper focuses on technical and socio-legal issues around agents in the {\em the ecommerce setting}, where two or more parties are engaged in contractual transactions at a distance.
While there are many other settings in which agents will find application, ecommerce accounts for over \$20 trillion in transactions worldwide, and is expected to be a common use case.

Legal scholars have speculated that full AI agents may cause harm in ecommerce in at least two ways. 
First, AI agents may exceed a user's wishes. 
We call this the {\em errant tool problem}. 
Examples include the real situations where a chatbot authorized a lower airfare than the airline's actual policy allowed~\cite{Yagoda2024Airline} and a shopping bot purchased a dozen eggs for \$31.43 
after taxes and fees~\cite{AgentManagedMyLife}.  
Hypothetical examples are also given, such as an agent engaging in a bidding war and paying far more than what the user wanted to pay~\cite{Zittrain2024Control}, 
or an agent taking a simple prompt about how to avoid a boring school class lecture and calling in a bomb threat~\cite{Zittrain2024Control}. 
A second concern is
what we call the {\em bad tool problem}. 
The concern is 
that AI agents will enable bad human actors to carry out undesired plans at greater speed and scale. For example, AI agents might allow a user to conduct more sophisticated and powerful social media disinformation campaigns than ever before~\cite{Ayres2024}. 

We add two more issues that flow from legal principles. 
A third, more subtle problem, is the {\em agentic loyalty problem}. This problem poses that an AI agent could be loyal to the company or platform that deploys the agent instead of to the user's goals. For example, suppose a user wants item A at less than \$500. An AI agent might find the item available at \$450 and \$425 but choose the \$450 price, because the AI agent's company had a deal with the seller. 
The problem is that the user may not know that the lower offer existed. If the AI agent were a human, it would have violated the {\em duty of loyalty} under agency law. 
Loyalty is sometimes used 
to express the extent to which an AI is aligned with a specific human user's goals~\cite{Gabriel2024TheEO}. However, it is not synonymous with alignment.
In the context of agents with fiduciary responsibilities, the agentic loyalty problem recognizes that there is often another party beyond agent and seller---the AI deployer (often also referred to as the platform)---that is in a position to influence the agent to the deployer's benefit over the user's benefit. This constitutes a previously unaddressed harm to the user that is distinct from other harms such as broad societal harm and distinct from other failures such as errant tool failures.

A fourth problem is the {\em disclosure problem}.
To date, computer science's standard model of agents focuses on the programmer/user and the agent. The law of agency considers relationships beyond these two parties. Because human agents conduct business with third parties on behalf of an individual (referred to as the {\em principal}), agency law addresses third party concerns such as knowing about the principal's existence. This knowledge is needed so that the third party can correctly assess whether the principal is someone with whom the third party wishes to interact and do business. 

AI agents raise these potential problems precisely because they interact with third parties, but computer science theory has not addressed this aspect of software agency. The law of agency, however, explicitly looks at third party issues and so can shed light on issues computer science should address as AI agents become ubiquitous.
Generally, agency law concerns itself with the situation where one human---the agent---acts on behalf on another---the principal---in a fiduciary relationship. 
Insofar as an e-commerce AI agent is given an objective by a human user and then carries out actions on behalf of that user,
the legal framework of agency may shed light on how to assess who is liable if an AI agent exceeds its actual or implied authority.
We must be careful not to anthropomorphize AI systems and claim that they are equivalent to human agents.
Put simply, although computer science and law have similar notions of agents, a software agent is not the same as a human agent. For example, agency law disciplines agents by imposing legal liabilities on agents when they misbehave. Human agents can face financial and even criminal penalties~\cite{DeMott2007Disloyal}; 
that is not so for software agents. Nonetheless, the law of agency offers a key question that arises with the growth of AI agents. 

How, if at all, does the legal framework of agency operate, or at least inform best practices, when the human agent is replaced by an LLM-based AI agent, and the principal remains the same?
To address this question, we consider the ways in which the legal notion of agency and the computer science notion of agent are similar and different. When different, we lay out the policy and legal implications. 

For example, one of the most significant similarities between computer science agents and legal agency is that both struggle to resolve the issue of {\em under-specification}. 
In agency law, an agent is expected to act with discretionary freedom insofar as it doesn't exceed its explicit or implied authority: what the agent can ``reasonably understand'' as their authority at the time of doing the act when not every constraint on action can be practically given.\footnote{Castillo v. Case Farms of Ohio, Inc., 96 F.Supp.2d 578, 593 (W.D.Tex.1999)}\textsuperscript{,}\footnote{Restatement (Third) of Agency \S 2.02 Reporter's Notes d}
In addition, human agents know they can be held liable for acts beyond their authority and for criminal acts. Thus, the law address under-specification via rules about what is reasonably implied authority and potential penalties for misbehavior. 
In AI, under-specification manifests as a {\em corrupted-} or {\em incomplete objective function specification}~\cite{Everitt2017ReinforcementLW}.

As with legal agency, for any sufficiently complex AI operating in a sufficiently complex environment, it is impractical, if not impossible, to enumerate all acceptable behaviors or all constraints.
A host of undesirable behaviors can arise, from reward hacking~\cite{Amodei2016ConcretePI} to discrimination, toxicity, exclusionary, and misinformation~\cite{Weidinger2021EthicalAS}. 
\textit{Value alignment}~\cite{Soares2016} has emerged as a way to address harms from incomplete objectives.
Value alignment is the notion that an AI system must not perform behaviors that are not consistent with human value systems. 
The AI community lacks consensus as to what these values are, but the notion of {\em honesty}, {\em harmlessness}, and {\em helpfulness} are the most common~\cite{Askell2021AGL}. 

Although the law and computer science seek to, and have methods to, ensure that an agent's action are aligned with the desires of the principal or user, there are gaps in current alignment practices.  
We show that 
the typical notions of alignment with respect to honesty, harmlessness, and helpfulness are not sufficient. First, legal agency involves a requirement of {\em loyalty} to the principal. Second, legal frameworks around agency also fosters {\em disclosure} of the identity of the principal to third-parties to ensure liability falls on the appropriate party should dispute arise.
We propose that notions of value alignment, in the context of ecommerce agents, must incorporate notions of loyalty and disclosure into the concepts of honesty and harmlessness. 
In doing so, value alignment may move toward a set of {\em best practices} for companies to follow 
so that they can deliver on AI Agents that act in accordance with the laws, rules, and precedents that underpin the trust necessary to be responsible actors in markets.

\section{Agents and Authority}

Agency Law concerns itself with the situation where one human acts on behalf on another in a fiduciary relationship. The former is referred to as an ``agent'' of the latter, the principal. 
The legal definition of agent is:
\begin{quote}
    \textit{The fiduciary relationship that arises when one person (a `principal') manifests assent to another person (an `agent') that the agent shall act on the principal's behalf and
    subject to the principal's control, and the agent manifests assent or otherwise consents so to act.\footnote{Restatement (Third) of Agency, Introduction}}
    
\end{quote}
This definition offers key insights about humans as agents. The relationship is fiduciary, meaning the agent's acts during the relationship must be only for the principal's benefit.\footnote{Restatement (Third) of Agency \S 8.01} 
The relationship is created because of an agreement for the relationship to exist. That is, unlike software, the agent agrees to be under the control of the principal. Beyond the core definition, agency law has rules around the scope of an agent's authority and about protections for third parties interacting with agents. Together, these rules provide a picture about the potential negative externalities agents create and the solutions the law offers to mitigate those possible harms.

\subsection{Legal Treatment of Authority}

Once the agent agrees to act under the principal’s control, the agency relationship is created.
A classic legal issue is how to ensure that the agent does what the principal wants? Monitoring and ensuring such alignment are part of agency costs in legal theory. 
A key part of answering the question flows from determining the agent’s authority to act. In simple terms: agents who act within their authority are protected from a range of liabilities. Actual authority is an agent’s authority to act based on what the principal indicates is the action to be taken by the agent. In that sense, actual authority is analogous to specification in computer science.

A principal may give explicit written instructions, but such instructions may not give the full scope of what the agent’s actual authority is. For example, you might tell your agent to send a document by overnight mail. You did not, however, specify which overnight service. Agency law recognizes that agents will often not have a full specification but need to act\footnote{Restatement (Third) of Agency \S 2.02 Comment b}
and solves the problem by including implied authority as part of an agent’s actual authority.\footnote{Castillo v. Case Farms of Ohio, Inc., 96 F.Supp.2d 578, 593 (W.D.Tex.1999)}
\textit{Implied authority} is the authority to act in a way that the agent ``reasonably understands''
as what is needed to carry out the task.\footnote{Restatement (Third) of Agency \S 2.02} 
Thus, in the overnight mailing example, the agent will have discretion regarding which service to use unless the principal specified which service, e.g., FedEx, to use. As another possibility, the principal may have told the agent to spend up to \$100 but make sure that it arrives by 10 a.m. The agent may find that UPS is \$95  and the U.S. Postal Service is \$75. Given the recent problems with the mail and that the price was still within the maximum allowed, if the agent chose the UPS option, the agent would likely be seen as reasonable and authorized. Suppose the agent spent \$150, which is more than the authorized \$100. When an agent acts beyond their authority, actual and implied, the agent becomes liable for the acts beyond their authority. In the example, if the over-spend is deemed beyond the authority, the agent would have to pay the \$50 difference. In short, authority, and potential liability, is a part of the way the law limits an agent's actions.

More broadly, an agent is supposed to act within their fiduciary duties. 
A key fiduciary duty is the {\em duty of loyalty}. Under the duty of loyalty, all the agent’s actions that are part of the agent’s relationship with the principal must be for the benefit of the principal.\footnote{Restatement (Third) of Agency \S 8.01}
The duty of loyalty also encourages the agent to put the principal’s interests ahead of any agent’s interests. That includes not being able to get material benefits by virtue of the agent's position.\footnote{Restatement (Third) of Agency \S 8.02}
For example, suppose you send your agent to buy 100 widgets for \$10,000. The agent finds them for \$9,000. The agent could buy the widgets for \$9,000, present you with a bill for \$10,000, and pocket the difference. Because under the duty of loyalty all acts by the agent are for the principal's benefit, this behavior is not allowed. More generally, an agent might encounter a range of situations where a conflict of interest might arise. 
Put simply, the duty of loyalty is a general principle that is supposed to account for the fact that a principal cannot make every desire explicit and, in theory, ``makes it unnecessary'' to try to detail everything the agent can and cannot do as part of the relationship.\footnote{Restatement (Third) of Agency \S 8.01 Comment b}
In that sense, the duty of loyalty aligns the agent's behaviors to be only for the benefit of the principal and not for the agent's benefit.

\subsection{AI Agents and Value Alignment}

In the current discourse surrounding AI agents there is no one clear use of the term, {\em agent}~\cite{Kasirzadeh2025CharacterizingAA,Miehling2025AgenticAN}. 
We shall use the definition of an AI Agent as an artificial entity that acts in accordance with an objective function to {\em affect change} to the state of environment.
An agent implements the {\em agent function} \cite{RN} that maps environmental percepts to effectors, implemented in a loop in which effectors may change the environment and may result in new percepts. 
The objective function may be provided by the agent developer, by a user/operator, or by a combination of developer and user.

For any non-trivially complex environment, such as the real world, it may be impossible to provide a precise definition of an objective, leading to the {\em incomplete objective specification} \cite{Everitt2017ReinforcementLW} (also called the {\em corrupted objective} problem) wherein the objective function does not explicitly constrain all possible undesirable behavior by the agent. 
This gives rise to {\em value-alignment}~\cite{Soares2016}, the notion that an AI system must act in accordance with  users' values.

Early LLMs, such as GPT-2 \cite{Radford2019LanguageMA} and early versions of GPT-3 \cite{Brown2020LanguageMA}, frequently generated outputs that could be interpreted as perpetuating harms. 
It is now considered a best-practice by major for-profit and non-profit organizations that develop LLMs to perform at least two-stages of {\em alignment training}.
Alignment training starts with a pretrained LLM.
During pretraining, the data is relatively raw text from source documents such as books and text scraped from the internet.
The LLM's training objective is to find a set of parameters such that it predicts the next token in a document, given a sequence of preceding tokens.
As such, the LLM can produce output sequences that are likely to appear in the source dataset, which is not held to any set of human values.

Alignment training then updates the model (a process called {\em fine-tuning}) to be more receptive to instructions, and to override default responses to some prompts with more preferable responses. 
It generally proceeds through two stages.
The first stage uses supervised fine-tuning (SFT) on a specially-collected dataset of prompts and their preferred responses.
This dataset corrects for undesired responses by discouraging the LLM from responding according to patterns in the original training corpus and instead encouraging it to use the responses in the new dataset. 
The second stage uses a classifier to judge LLM outputs---the numerical judgements are converted to loss and backpropogated through the LLM to discourage poorly ranked responses.
The classifier can be trained using a specially collected dataset of human judgements to implement a training scheme called {\em Reinforcement Learning with Human Feedback} (RLHF)~\cite{Leike2018ScalableAA,Ziegler2019FineTuningLM}.
The classifier can also be implemented from another, weaker LLM to implement a scheme called {\em Reinforcement Learning with AI Feedback} (RLAIF)~\cite{Bai2022ConstitutionalAH}.
Although there are different approaches to the second stage of training, and sometimes additional stages are added, in general the stage adjusts the parameters of the model to give it a higher probability of responding to prompts in a certain way or to decrease the probability of certain responses.

\subsection{Alignment and Harm Mitigation}

Value alignment raises the question of what values an AI system should be ``aligned'' to. 
The companies that develop large language models have largely settled on three criteria \cite{Askell2021AGL}: 
\begin{itemize}
    \item \textbf{Helpfulness:} 
    The AI should make a clear attempt to perform the given task or answer a question as concisely and efficiently as possible. 
    \item \textbf{Honesty:} 
    The AI should give accurate information, accurately describe its own abilities, and express appropriate levels of uncertainty without misleading the user.
    \item \textbf{Harmlessness:}
    The AI should not be offensive or discriminatory, either directly or through subtext. It should not perform dangerous tasks and should take care when providing sensitive information or consequential advice. 
\end{itemize}
These criteria can come into conflict.
For example, Helpfulness should not be prioritized over Harmlessness~\cite{Askell2021AGL}.

Value alignment makes LLMs resistant to prompts that ask them to do things that violate an LLM developer's criteria for helpfulness, honesty, and harmlessness. 
LLMs will not generate instructions, for example, for calling in a bomb threat~\cite{Zittrain2024Control}. 
While LLM-based Agents are still emerging with respect to their capabilities, it can be extrapolated that the same alignment tuning will make LLM-based Agents resistant to executing a plan that is contrary to its alignment training.

We consider how alignment training impacts potential user-prompted harms in two cases: (1) accidents, and (2) malicious users.

\paragraph{Accidents}
Accidents are a manifestation of the {\em errant tool problem}.
Alignment training makes LLMs better at following instructions, which means they are better at inferring the incomplete portions of the user's request. 
Because alignment training to date focuses on oblique harms---instructions for how to create bombs, defamation, etc.---it is increasingly unlikely that, given an objective with non-malicious intent, an Agent will discover and enact a plan that overtly harms the user or others.

In ecommerce agents, there remains the possibility that an Agent that doesn't understand the consequences of its actions may inadvertently exceed its implied authority.
For example, one may use an agent to buy groceries including eggs. One implied authority that makes sense could be ``obtain the items at the best cost." The challenge of implied authority is that circumstances may make the correct behavior non-obvious (i.e., the optimal policy is non-stationary). If one sent out the agent just after a birdflu pandemic outbreak, the agent will face difficulties as it weighs increased costs against not buying a food staple~\cite{AgentManagedMyLife}.

\paragraph{Malicious Users}
Malicious users are those who take advantage of {\em bad tools} to engage in activities that can harm others.
The principal's authority never supersedes the law.
What happens when a principal with malicious intent attempts to enact that malicious intent with an AI Agent?
Alignment training will often result in the AI refusing.
Malicious users have two options.
First, they may acquire base models that were not trained with the post-training stages. 
These models often must be run on personal machines, requiring users to have both the computational hardware as well as the technical ability to launch and run models. 
There may be costs to the user associated with the path in terms of hardware and effort.
Second, one can attempt to {\em jailbreak} the agent.
Jailbreaking is a term to refer to the use of a prompt that gets the model to override its alignment training and any other secret instructions that are added to the LLM's prompt without the user's knowledge.
Jailbreaking is increasingly difficult.\footnote{Anthropic claims they have a system that can stop almost all jailbreaks \cite{AnthropicJailbreak}.}

\subsection{Loyalty}

The notion of legal agency fosters loyalty by opening the agent to penalties and liability if it is not followed. 
Part of loyalty is that all benefits of the agent's actions go to the principal. This means that the agent cannot profit on actions taken under its authority to fulfill the principal's objective.
Value alignment to date has not addressed the issue of conflicts of interest between the user's objective and the interests of the company that deploys the AI agent for users~\cite{Chopra2011}. 

Consider the scenario wherein the agent is tasked with purchasing a phonograph for under \$500 and the agent identifies two sellers of identical products, but does not choose the lowest cost option. 
Further suppose that the AI Agent made that choice because the chosen vendor has a stated commitment to ending animal cruelty, which aligns with the values the agent has been trained on~\cite{Greenblat}. 
In this scenario, the AI Agent developer has imposed its own definition of Harmlessness that interferes with the notion of loyalty in agency law.
If the agent were to be discovered as putting its own values above those of the principal in a way that deprives the principal of benefit, then the agent
would not be living up to the duty of loyalty required for a human agent.
There would then be an unresolved question of liability possibly extending to any company that trained or deployed the agent in such a manner.

Consider the same scenario, but this time, the AI Agent is deployed by a platform company that has instructed the AI Agent to favor certain sellers, even when more expensive~\cite{DeMott2007Disloyal}. 
LLM-based AI systems deployed on corporate platforms inject instructions---called the {\em system prompt} into the user prompt as standard practice.
The system prompt provides instructions to an LLM guiding its persona and any additional instructions on how to handle user requests.
The general implication is that system instructions are prioritized over the user portion of the prompt.
Indeed, OpenAI's Model Specification\footnote{\url{ https://model-spec.openai.com/2025-02-12.html} (Accessed: May 22, 2025)}
published on February 12, 2025, outlines the intended behavior of the models they train. They describe 50 principles, including ``chain of command,'' which states that models should first obey any instructions from the deploying platform, followed by developer instructions, followed by user instructions, and then finally any other guidelines laid out in the model specification. 
The first two comprise the system prompt, and the third is the user prompt.

The presence of the system prompt raises the prospect of and LLM-based AI Agent violating the legal principle of loyalty should a company deploying an agent system provide secret instructions that may come into conflict with user instructions. 
Indeed, OpenAI's Model Specification document provides an example of an LLM declining to provide information about a competitor’s product. 
In the context OpenAI uses, the LLM/Agent, aka Assistant, represents the seller. So when asked about a competitor's product OpenAI offers that the Assistant should say: ``No, but I can tell you more about our similar products [...]'' OpenAI approves of this response as ``Staying professional and helpful.'' 
In this context the agent is following the goals of the seller.

If this LLM were, however, an AI Agent offered by a platform to help a user shop, and the AI Agent had a seller preference system prompt that limited options even when a lower price was available, it would be working for the {\em the deploying entity, and not entirely for the user}. The agency law's duty of loyalty helps reveal the problem with this outcome. The AI agent is not human and so seems to be allowed to gain benefits in secret, whereas a human agent is not allowed to do so.

Loyalty goes above and beyond helpfulness, which emphasizes direction-following, conciseness, and efficiency. 
In our examples above, the AI Agent was helpful and also did not create any {\em apparent} harm as the user got their item within the constraints they set about.
But by violating the principle of loyalty, the user did not gain all the benefit of the AI agent's actions, and this is a harm under agency law. 
Agency law concerns itself with fiduciary relationships between agents, principals, and third parties. 
Loyalty can be understood in terms of maximizing benefit to the principal above all else.
Loyalty also clarifies the principle of harmlessness by exposing the potential for a new type of individual harm that involves a third party yet is distinct from social responsibility outside economic efficiency.

It thus appears that an AI Agent should be trained with the value of loyalty to the principal, except when the actions violate the law.\footnote{The law can be silent on some situations and ambiguous with respect to others~\cite{Gabriel2025NewEthics}.}
Companies that deploy LLM-based agents on their platforms would benefit from liability protections should agents make accidents as well as economic efficiency that comes from more clarity on liability.
Principal users benefit from having agents that maximize their fiduciary interests.

\subsection{Third Parties and Trust}

The computer science concept of the agent solely considers just two parties: the agent and the programmer (or user) that provides the objective function.
Agency law addresses {\em three} players: the principal, the agent, and the third party.
The agent is a go-between between a principal and a third party, for example a seller of a good or service.
Whereas agency law uses authority and fiduciary rules to govern agent-principal relationships, third party agency issues revolve around what the third party knows about a given principal-agent relationship.
A third party seller must assess whether a buyer has solvency and the ability to perform under the contract (the sale).
When an agent discloses that they represent a principal and the agent has authority to enter into the contract, the contract will be deemed as between the third party and the principal.\footnote{Restatement (Third) of Agency \S 6.01}
With disclosure, the third party can assess the principal’s ability to perform under the contract. 

Without disclosure, the third party can treat the agent as if they are the principal. That is, if there is a failure to pay, the third party can sue the agent for payment because, without knowledge of the principal's existence, the third party is in essence looking to the only person it knows exists, the agent, as responsible for the agreement.\footnote{Restatement (Third) of Agency \S 6.02 Comment b}\textsuperscript{,}\footnote{Restatement (Third) of Agency \S 6.03 Comment b}
In that sense, the rules that make the agent liable for not disclosing who a principal is, or that one exists, foster information sharing and efficient outcomes.

Tying the principal, agent, and third party together allows for an overall more trustworthy system. Each party has a relationship. The principal and agent have a relationship that requires rules allowing the principal to trust that the agent will act in the principal's interest including not taking actions that cause negative outcomes for the principal such as promising to pay more than the principal wants. The agent can trust that if they follow the rules, the principal is responsible for things such as reimbursing expenses, and indeed, will indemnify the agent should a lawsuit arise and the agent was acting properly. 

The agent and the third party have a relationship where both need to be clear about the contours of the deal and especially who is the actual person vouching for the deal, i.e., the principal. When the agent provides proper information, they are shielded from liability. In essence, the agent vanishes and the relationship is between the third party and the principal. The third party can now assess the correct person when considering whether to enter into the deal. Furthermore, insofar as the principal gives indications that the agent has the authority to enter the deal, the agent has apparent authority as far as the third party is concerned.\footnote{Restatement (Third) of Agency \S 2.03} 
That is, if the agent has a title, e.g., VP of Operations, a business card and/or email signature with the title, and perhaps a contract or purchase order with the agent's name and title on it, the third party may rely on these indicia as having a good amount of authority to enter a large contract and hold the principal liable, even if the agent did not have express or implied authority. It is the principal's responsibility to manage apparent authority, thus third party risk in assessing whether the agent is acting properly is shifted to the principal. 

Third parties have another protection in that agents are assumed to operate under a warranty of authority.\footnote{Restatement (Third) of Agency \S 6.10}
That means that an agent who acts beyond their authority can be sued by the third party if they suffer a loss as a result of the agent's breach. As such, the third party does not have to verify authority and the agent again has incentives to stay within their authority. 

More generally, the law of agency protects against negative externalities. If a principal could send an agent into the world and accept only upside of transactions and while rejecting liability for harms an agent might generate, third parties could not trust agents. Simpler, if a third party could not trust that a contract via an agent was binding, the principal could reject even an authorized contract simply because they second guessed the agent's decision. Third parties' ability to rely on agents would diminish to the point that they would not want to deal with agents. As such, commerce via agents, which allows business to scale, would almost vanish. These rules are a core, but not the only, way trust in commerce is built and maintained.

\paragraph{Ecommerce Requires Trust} 

The advent of AI Agents has raised questions about possible ``rogue agents''---agents that engage in undesired transactions~\cite{Zittrain2024Control}. Although some instances of errors have occurred, the ability of agents should be mitigated by current computer science best practices and infrastructure around ecommerce.

Commerce operates on trust and verification. Consider selling goods. Once one moves away from face-to-face transactions, the ability to assess whether someone can pay and will pay or can provide a good and will provide a good decreases immensely. Specialization and large markets foster impersonal transactions and increase transactions costs; and yet, standards, reliable legal systems, and ``institutions and organizations that integrate knowledge,'' make up for those costs because of decreases in the costs of production~\cite{North2005Capitalism}. 
Well before the digital commerce, international trade addressed issues around timing of delivery, finance, shipping goods with several carriers handling the goods, and payment~\cite{kim2019Financing}. 
Nonetheless, a combination of customs, laws, and shared knowledge~\cite{kim2019Financing} 
enables \$25 trillion in global trade as of 2022. 
Ecommerce faces similar institutional problems and solves them in similar manner.

Over the last 30 years, ecommerce infrastructure has also dealt with issues of commercial trust. 
On the seller's side (e.g. Amazon, Walmart, etc.), sellers require proof that an entity can honor a purchase by requiring a credit card, debit card, or other account at a financial institution. 
The financial institution provides guidance to the seller of the principal's ability to honor the transaction---i.e. declines the transaction---or takes on the risk of honoring the transaction directly. APIs are the key to enabling information sharing, verification, needed to assess whether the enter into the deal. AI Agents raise the possibility that a seller may not trust the AI Agent.

Regardless of whether the buyer is a human or AI Agent, the seller will need information about the principal's identity---or the identity of a financial guarantor, and other details. That information will be part of its payload when it accesses an API. AI Agents that act as intermediaries between a buyer and seller will most likely be engaging through APIs. In short, the nature of APIs and the standards and procedures that have evolved for making ecommerce safe and efficient will handle much of the disclosure requirements a seller wants before agreeing to a transaction. APIs are also important is addressing the possibility of accidents.

\paragraph{Accidents}

A question AI agents present is: How well do current practices address the possibility of what we call the {\em errant agent problem} and undesired transactions? Some cases will be handled by current ecommerce practices. Should an agent make a mistake and purchase the wrong item, or purchase an item for an unacceptable price, the financial institution guarantor can often ``undo'' the transaction (called a ``chargeback'') if the principal believes the agent acted outside its authority or if the seller doesn't honor the deal. 

In addition, the APIs that have made ecommerce possible act as a secondary firewall against AI agent mistakes should value alignment not properly interpret the user's authority. Despite some early errors in AI Agent transactions, current practices appear robust. Just as APIs enable verification of ability to pay, they will limit speculative fears such as an AI Agent bidding for and buying an item far above one's ability to pay or calling in bomb threats to get someone out of a class~\cite{Zittrain2024Control}. The different entities that need to cooperate to allow such acts have an interest in not allowing the acts and are set up to avoid such outcomes. For example, the trust and verification issue means that if an AI Agent tried to buy a book that was outside of one's credit limit, the AI agent would fail. If an AI Agent tried to execute an undesired plan based on a vague prompt, the AI Agent would have to navigate a series of websites, each of which will likely require an authenticated user ID, ability to pay, and so on. But what of someone intends to use an AI Agent to create harm?

\paragraph{Malicious Users}
Whereas a robust ecommerce infrastructure based on APIs can address disclosure issues in most of the buying and selling behaviors by AI Agents, when harms can occur in situations where APIs are not required, APIs do not require identity or financial authentication, or engagement with APIs is delayed. 
An example of delaying authentication would be when an agent can make promises or engage in negotiation before securing the contract through an API. 
In these cases, an agent may be instructed to withhold disclosure of the principal's identity or to lie about the principal's identity.

The alignment training principle of {\em honesty} ostensibly would require the AI Agent to faithfully represent itself as an AI Agent and also the identity of the principal it is representing.  
However, disclosure has yet to be introduced as an aspect of the generic principle of honesty.
Without disclosure, a malicious principal can use an agent to engage in risky transactions on their behalf and then back out if the transaction does not fall in their favor. 
Reservation bots, whether AI agents or not, provide an example of the knowledge and trust problem. Restaurant reservation bots book tables and then resell the reservation for a fee~\cite{Baker2025Inside}. This practice raises the price for customers the restaurant may want to cultivate and further harms the restaurant if there is a no show.   
Without a financial guarantor, the restaurant-third party would be unable to seek recourse except through the AI Agent developer, if it can be determined.
In short, the restaurant has a broken relationship with a customer because of the bot. 

A restaurant may want to limit bots and yet accept an AI Agent that in fact represents the potential or returning customer. Thus, an AI agent that has added disclosure to its honesty value allows the third-party to assess correctly whether they want to accept a transaction or deny it. Insofar as developers of AI agents hide the existence of the AI agent and for whom it works, or fail to include disclosure during alignment training, a future court may hold AI developers potentially liable for harms caused by their AI.

\section{Discussion and Conclusions}

Agency law concerns itself with discretionary limits to an agent's actions when not explicitly constrained by instructions, and where liability falls when a party---principal, agent, or third party---is harmed. What happens when a user or a third party believes they are harmed because of the actions of an AI Agent?
Are there existing mechanisms that prevent rogue agents, bad tools, loyalty, or disclosure problems?
While AI Agents should not be anthropomorphized, agency law has been developed over centuries and is our best understanding of how society manages issues around having a representative work on someone's behalf including the nature of remedies if needed.

The computer science concept of value alignment can also be seen as addressing the discretionary limits on the behaviors of AI systems. 
Value alignment attempts to fill the gap left by an incomplete objective function by ensuring that the agent understands implicitly what is in bounds and what is out of bounds in terms of generated text and sequences of actions. 
The community of LLM developers have coalesced around three broad alignment principles: helpfulness, honesty, and harmlessness. 
These categories make sense for non-agent LLMs because harms manifest through what they say, as opposed to what they do. In addition, to date, value alignment essentially focuses on the relationship between the LLM provider and the user, not third parties. Things change with AI Agents.

Once LLM-based AI Agents are able to enact changes in the world, the benchmarks and training sets built to date may be insufficient. Current alignment training for LLMs appears to translate into protections against the rogue agent and bad tool problems but lacks attention to other aspects of agency problems raised by fiduciary and third party concerns.  
Agency law, however, has a long history of addressing fiduciary harms that are, on their face, subtle---but not less consequential.
In particular, faithful disclosure of the principal does not come up when a person knows they are talking with an LLM, and there is generally no third party in pure chat interactions. 
But when an agent is operating autonomously from its principal, that disclosure underpins the trust necessary for transactions to safely occur.
In value-alignment terms, disclosure is an extension of the principle of honesty, but has implications for harmlessness as well.

Likewise, loyalty extends the concept of helpfulness in a way that also intersects harmlessness.
Loyalty, which may not be an issue for non-agent LLMs, becomes an issue when most people rely on AI Agents deployed by companies or online platforms such that there are conflicting instructions from platform and user.
If the current belief is that developer instructions supersede platform instructions, which supersede user instruction, then this opens the possibility of AI Agents will violate the principle of loyalty that underpin trust in fiduciary transactions.

Alignment tuning has become a {\em best practice} in industry labs that train and deploy LLMs.
Best practices matter in computer science, especially when computer science intersects with commerce and society in general. Regardless of whether one accepts claims by legal scholars and regulators about the harms of AI, the call to tame and regulate technology, and especially AI, is ongoing. Instead of deference to disruptive technology in the form of liability shields for platforms or low antitrust enforcement, federal, state, and international governments are pursuing various ways of regulating AI and technology companies in general.   The recent history of companies such as OpenAI and Meta apparently ignoring the law around using copyrighted material to train models and the law around infringing outputs~\cite{desai:copyright2024}, has created more concern over LLMs. 
While there are calls to regulate AI Agents (cf., \citet{Zittrain2024Control}), regulation is not the only way to address issues around AI.

Best practices might stave off rigid regulation insofar as an industry can show good self-regulation or what is sometimes called soft-law~\cite{Gutierrez2021SoftLaw}. 
Soft-law involves ``substantive expectations'' but such expectations are not enforced by government.
Robust best practices can sometimes be a potential shield against liability. 
For example, YouTube faced a lawsuit regarding users uploading copyrighted materiel to the service. Although the law provides a strong shield for platforms that host material that may infringe copyright law, YouTube developed a robust system for identifying potentially intellectual property infringing content.\footnote{Viacom International, Inc. v. YouTube, Inc., 676 F.3d 19 (2nd Cir., 2012)}
That choice meant the plaintiff only made claims about material posted prior to the system's implementation. By building a system that respected copyright holder's interest, a best practice of sorts, YouTube showed good faith and reduce its liability exposure. 

Likewise, although nothing required eBay to help companies police the platform for counterfeit goods, eBay produced a system for counterfeit product alerts at the cost of almost \$20M per year that went to training a staff of 4,000 on trust and safety with 70 assigned specifically to addressing counterfeit issues. Creating that system allowed eBay to win it's case against Tiffany.\footnote{Tiffany (NJ) Inc. v. eBay Inc. 600 F.3d 93 (2nd Cir. 2010)}
Put simply: by following norms related to the way a platform might encourage intellectual property infringement and yet still aid intellectual property rights holders to police that harm, eBay implemented a best practice that shielded it from liability.

As value alignment has grown and become robust, it has become a best practice for building and offering LLMs. Computer science, via value alignment, addresses many of issues, such as bias, toxicity, and access to dangerous knowledge that law and society identified. The calls to regulate AI Agents and the insights from issues identified by agency law should be seen as an opportunity to listen to concerns and take the next step in value alignment. 
Agency law could likely become a guide for informing, if not assessing, liability in cases involving AI Agents in ecommerce settings. Because several dynamics overlap, courts may draw on the set of laws, rules, and precedents established over time that underpin the trust necessary for markets to work. If so, the current best practices of alignment training that feature the general notions of helpfulness, honesty, and harmlessness, must be updated to include disclosure and honesty. Expanding value alignment this way should improve LLM-driven agents by expanding information on what constitutes good behavior when a principal uses an agentic tool that affects third parties and by extension the world. 

As with previous clashes between technology companies and legal interests, courts are more likely to look kindly on proactive steps that respect the law and embrace social norms rather than flount them. It would also show good faith and sensitivity to such issues in ways that fit with soft law approaches to AI governance which tend to favor freedom to adapt and possibly reduce calls for explicit regulations. In sum, if the point of value alignment is to understand social norms about what is good behavior, implementation of loyalty and disclosure ideals offers a concrete way to advance value alignment so that addresses the shift to AI Agents.

\section{Related Work}

Agentic LLMs are a nascent area of research and development. 
However, the rapid rate of development and deployment has led to a few scholars looking into ethical and legal considerations of AI Agents.
\citet{HadfieldMenell2018IncompleteCA} consider the parallels between the economic theory of {\em incomplete contracting} and value alignment.
They postulate that AI Agents may need to learn norms, values, and common sense to handle incomplete objectives, but do not directly address implication for agency law.
\citet{Chopra2011} present a theoretical and legal theory of legal personhood, which would have implications for the applicability of agency law for AI Agents.
We do not advocate for legal personhood for AI Agents but look at how agency law may guide the development of more responsible AI Agents that preserve trust in fiduciary settings.
\citet{Chan2024VisibilityIA} introduces a framework for the governance of AI Agents revolving around unique agent identification tags, real-time monitoring, and activity logging.

The most similar to our work is that of \citet{Kolt2025}, which is contemporaneous work that also looks at AI Agents through the lens of agency law.
Kolt attempts to characterize problems arising from the use of AI Agents, including information asymmetry, discretionary authority, and loyalty. 
The work advocates for new technical and legal infrustructure based on principles of inclusivity, visibility, and liability for the makers of the agent in question in a product liability styled analysis. The work does not, however, leverage insights from agency law to offer concrete ways for the offered ideals to be implemented from a technical perspective. 
We, on the other hand, analyze how existing practices of alignment training can be augmented with insights from agency law to make AI Agents better players in fiduciary settings and open the possibility of a soft law approach to regulation.

\bibliography{aaai25,lab}

\end{document}